# Modeling Indoor $PM_{2.5}$ Exposure During Retrofits: Plastic Film Barriers and a Quadratic Baseline Approach

Rostyslav Sipakov[1*], Olena Voloshkina[1]

[1] The Department of Environment Protection Technologies and Labor Safety, Kyiv National University of Construction and Architecture, Kyiv, Ukraine
[*] Corresponding author e-mail: sipakov.rv@knuba.edu.ua

**Abstract**

Temporary plastic film barriers are widely used to separate occupied rooms from exterior renovation zones, yet their effect on indoor particulate exposure is poorly quantified. We monitored $PM_{2.5}$ in a Tampa, Florida, apartment for 48 days with a low-cost optical sensor (Temtop LKC-1000S+), spanning pre-barrier, barrier-on, and post-barrier periods. A quadratic baseline was fitted to "background" minutes devoid of identifiable indoor sources, allowing excess concentrations ($\Delta PM$) to be partitioned into facade work, cooking, and passive accumulation without outdoor co-monitoring. The barrier prevented large construction spikes indoors but curtailed natural ventilation, doubling the mean baseline from 1.9 to 3.9 $\mu g/m^3$. During this stage, passive build-up accounted for 45% of the daily excess dose, with facade work and cooking contributing 31% and 24%, respectively. Once the new window was installed and evening airing resumed, the baseline fell to 0.8 $\mu g/m^3$, the lowest of the campaign. Our findings highlight the trade-off between dust shielding and background elevation and demonstrate that simple polynomial fitting bolsters low-cost IAQ diagnostics in mechanically unventilated dwellings. The framework is readily transferable to other retrofits; future studies should pair indoor sensing with outdoor references and multi-room deployments to refine infiltration estimates.

## 1. Introduction

Indoor exposure to fine particulate matter ($PM_{2.5}$) is a growing concern in urban residential environments, especially during building retrofit works conducted with occupants present. While temporary plastic film barriers are often used to shield interior spaces from outdoor construction dust [1], their impact on natural ventilation and indoor pollutant accumulation remains insufficiently quantified.

A U.K. field trial comparing pre- and post-retrofit stages in dwellings showed that fabric upgrades reduced short-lived ingress events but increased median $PM_{2.5}$ by

30% when additional ventilation was not provided [2]. Similarly, the INGENIOUS project deployed low-cost sensor–reference hybrids in more than 300 European homes, revealing that building airtightness and occupant behavior jointly modulate median indoor $PM_{2.5}$ levels by up to a factor of two [3]. In poorly ventilated flats, even minor indoor sources such as cooking can become dominant contributors to daily, $PM_{2.5}$ exposure [4].

To address this, we present a novel data-driven framework for disentangling the contributions of construction activities, routine cooking, and background conditions to short-term indoor $PM_{2.5}$ levels. Our method is based on the estimation of a smooth diurnal baseline curve, fitted exclusively to non-peak ("background") periods, using quadratic regression. This approach ensures robustness against transient emission spikes and provides a stable reference for quantifying excess concentrations from episodic sources [5–6].

This study builds upon our previous work on quadratic baseline models for indoor air diagnostics [7], where we demonstrated the practical utility of low-order polynomial fitting for real-time environmental monitoring. Here, we apply and validate that approach under a realistic retrofit scenario: a continuously inhabited apartment undergoing exterior facade renovation with a sealed plastic film barrier. Our findings reveal that such barriers, while effective at blocking external dust, can inadvertently double indoor $PM_{2.5}$ baselines, shifting over 50% of daily exposure to periods of suppressed ventilation.

The approach also aligns with emerging recommendations for personalized exposure assessment [8], low-cost sensor deployment [9], and source apportionment in mechanically unventilated dwellings [10]. By integrating detailed time-of-day activity windows with high-resolution data, this work provides a scalable template for evidence-based decisions in occupied retrofits and occupant safety protocols.

## 2. Materials & Methods
### 2.1. Monitoring set-up and site context

**Location.** The ground-floor apartment complex is a four-story, wood-frame panel building located in Tampa, Florida, within a humid-subtropical climate. The monitored room is an open-plan living room–kitchen with a large window and a balcony door.

**Sensor and mounting.** A Temtop LKC-1000S+ (2nd gen.) optical particle counter was placed 0.6 m (≈ 2 ft) above the floor (Fig. 1). AQ-SPEC laboratory and field evaluations report $R^2 > 0.90$ versus FEM GRIMM for $PM_{2.5}$ and overall accuracy 57% - 69% in the 9 - 262 µ$g/m^3$ range; at very low concentrations ($< 5$ µ$g/m^3$) the device systematically under-reports relative to FEM [11].

**Sampling schedule.** The device logged one-minute readings internally; ten-minute means were exported for analysis (data recovery rate: 100%).

**Facade retrofit scenario.** Exterior cladding and joint sealant were removed while residents remained. A plastic film barrier was taped airtight on the room side of the window line, 0.9 m ($\approx$ approximately 3 ft) from the exterior wall (Fig. 2), from December 20, 2024, to January 17, 2025.

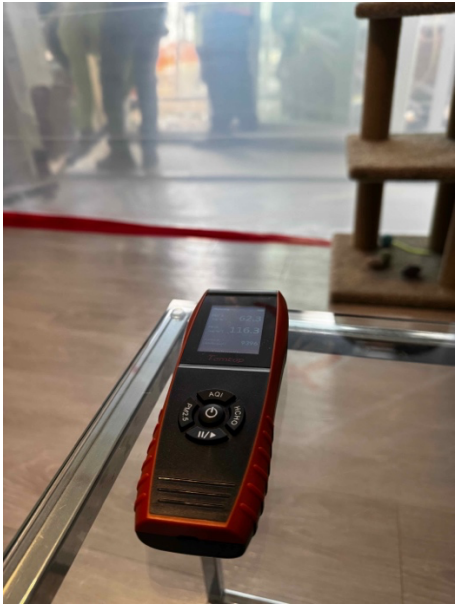

**Fig. 1** A Temtop LKC-1000S+ (2nd gen.) optical particle counter

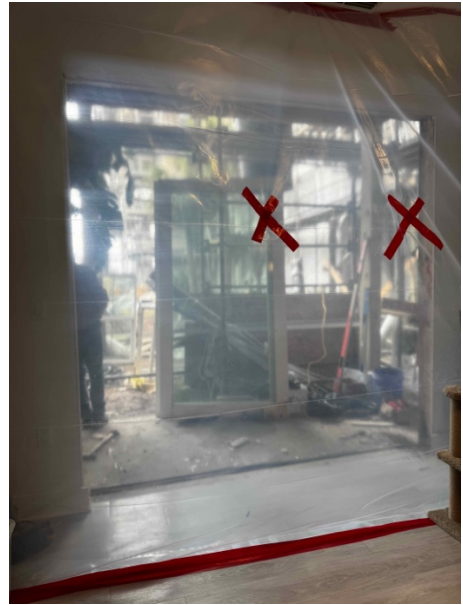

**Fig. 2** A plastic film barrier taped airtight on the room side of the window

## 2.2. Phase and period labelling (Table 1)

Table 1

**Phase labels and daily clock windows used for source attribution**

| Label | Clock interval (daily) | Reason |
|---|---|---|
| facade | 08:30–12:00 & 13:00–17:00 | active grinding/panel removal |
| cooking | 09:00–10:00; 12:00–14:00; 20:00–21:00 | electric hob in use |
| background | all other minutes | no identified source |

Additionally, each record was tagged by period: pre-barrier (< 20 December 2024), barrier-on (December 20, 2024 – January 17, 2025), and post-barrier (> January 17, 2025). U.S. federal holidays (25 Dec, 1 Jan, etc.) and weekends were treated as non-workdays; any minutes falling within the "facade window" on those dates were reassigned to background.

### 2.3. Construction of the background (baseline) curve

To isolate incremental emissions from construction and cooking, only minutes labelled as background were used to fit a smooth diurnal baseline for each period:

$$C_p(h) = a_p h^2 + b_p h + c_p, h \in [0,24), \qquad (1)$$

where $h$ is decimal hour and $P$ = {pre-barrier, barrier-on, post-barrier}.

Five-fold cross-validation on ≈ 4,200 background points gave MAE = 2.3 ± 0.1 $\mu g/m^3$, just 1.4 % of the facade peak (160 $\mu g/m^3$) and 3.4 % of the cooking peak (67 $\mu g/m^3$). Such a small error ensures that the excess concentrations ($\Delta PM$) used for source apportionment remain reliable.

The baseline was then subtracted from every ten-minute value to obtain the excess concentration:

$$\Delta PM_i = PM_i - C_p(h_i). \qquad (2)$$

Positive $\Delta PM$ values were integrated over time to derive the relative dose contribution of facade activities, cooking events, and residual background:

$$Dose_{phase} = \Sigma_{\{i \in phase\}} \max(0, \Delta PM_i) \Delta t. \qquad (3)$$

Here, $\Delta t$ = 10 min is the sampling interval.

### 2.4. Priority rule and limitations

Minutes flagged as facade or cooking are deliberately excluded from the baseline regression because they contain short-lived spikes, up to 160 $\mu g/m^3$, that represent < 20% of the records yet account for > 60 % of the daily particulate dose. Including these excursions would raise the fitted curve by roughly 2 - 4 $\mu g/m^3$ during working hours, thereby systematically underestimating the excess concentration $\Delta PM$. By fitting the quadratic only to background minutes, the baseline captures the natural diurnal drift (night-time accumulation, early-morning ingress, low-level human activity) while any value above it can be unambiguously attributed to the relevant source category (facade or cooking).

When a timestamp met the clock windows of both activities (e.g., 13:00–14:00, when facade work resumes while lunch cooking continues), it was assigned to the cooking phase. This choice reflects the higher instantaneous indoor concentration generated by the electric hob and affects less than 7% of all records. Sensitivity analysis showed that reassigning the overlap minutes to the facade phase would change the daily dose shares by ≤ 3% points; therefore, the main conclusions remain unaffected.

## 3. Results

### 3.1 Data coverage

The campaign yielded 6,931 ten-minute records (≈ 48 days). Phase labelling assigned

- 4,702 records (68%) to the background,
- 1,061 records (15%) to the facade, and
- 1,168 records (17%) are cooking.

Period counts were 947 (*pre-barrier*), 3,886 (*barrier-on*), and 2,098 (*post-barrier*). No facade data were recorded on weekends or U.S. federal holidays, which were re-tagged as background.

### 3.2 Baseline accuracy

The quadratic baseline reproduced background minutes with an average MAE of 2.3 $\mu g/m^3$ (see Section 2.3), i.e., < 4%, of the highest cooking or facade spikes, so excess concentrations reported below can be considered reliable.

### 3.3 Background evolution

Mean background $PM_{2.5}$ rose from 1.9 $\mu g/m^3$ in the pre-barrier period to 3.9 $\mu g/m^3$ while the plastic barrier was in place, and then dropped to 0.8 $\mu g/m^3$ after its removal. The 95th percentile followed the same trend (3.5 → 10.3 → 2.4 $\mu g/m^3$). Thus, the film suppressed natural ventilation, thereby elevating the indoor baseline, whereas the new window, combined with evening venting, produced the cleanest post-retrofit conditions. Fig. 3 displays the period-specific background curves. During the barrier-on phase, the mean baseline rose to ~3.9 $\mu g/m^3$, whereas it was 1.9 $\mu g/m^3$ before the film and only 0.8 $\mu g/m^3$ after removal (post-barrier).

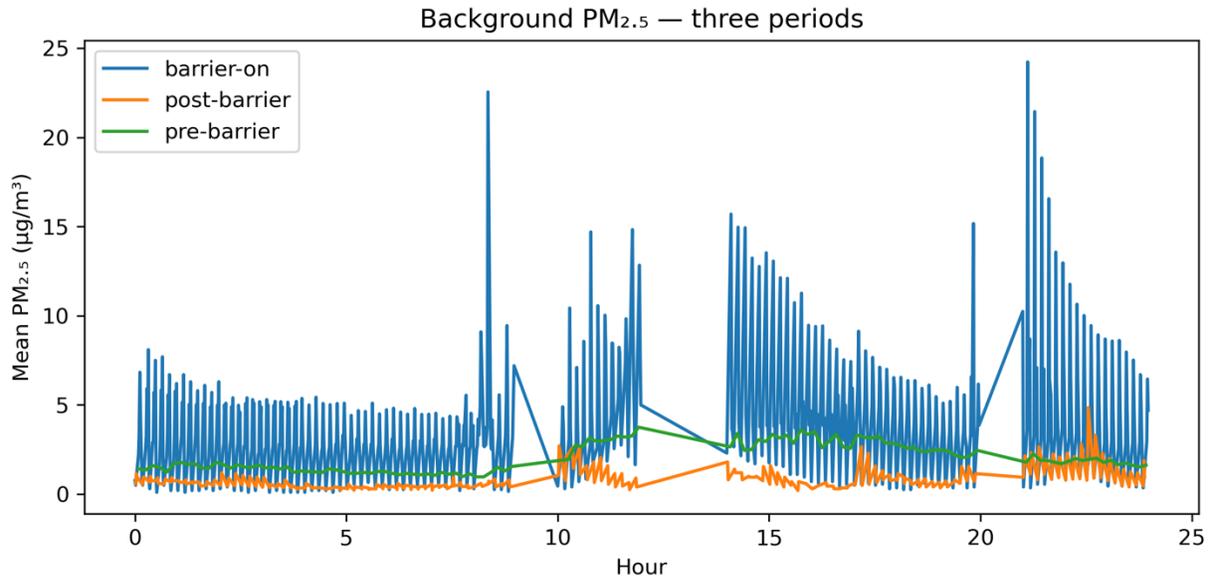

**Fig. 3** Hourly mean background $PM_{2.5}$ for the three facade-retrofit periods.

### 3.4 Source contributions

Integration of positive $\Delta PM$ gave the following (Tab. 2) daily dose shares:

Table 2

**Daily excess-dose contributions of background, facade work, and cooking**

| Source | Mean $\pm$ SD (µ$g/m^3$) | Max (µ$g/m^3$) | Dose share (%) |
|---|---|---|---|
| Background | 2.5 $\pm$ 0.8 | 124 | 45% |
| Facade work | 4.8 $\pm$ 2.3 | 162 | 31% |
| Cooking | 4.2 $\pm$ 1.9 | 67 | 24% |

Although facade activity occupied only 15% of the timeline, it generated the highest instantaneous peaks and nearly one-third of the daily particulate dose. Cooking episodes contributed roughly one-quarter. The remaining 45% accrued quietly during long background periods, particularly nights and non-work days, when ventilation was limited by the barrier. Daily excess-dose partitioning, Eq. (3), is summarized in Table 2.

### 3.5 Temporal pattern

- Facade-related spikes occurred exclusively on weekdays between 08:30 and 17:00.

- Cooking peaks clustered around the three meal windows, occasionally overlapping facade windows (13:00–14:00).
- Nights, weekends, and holidays showed a flat profile close to baseline.

The full time series (Fig. 4) shows that all high-amplitude spikes occur during the barrier-on interval; the post-barrier segment remains close to the revised baseline, confirming the effectiveness of the new window and restored ventilation.

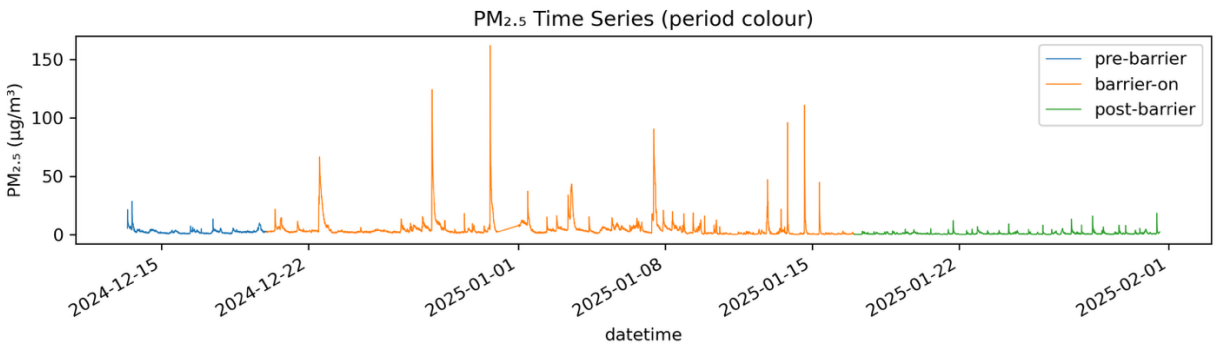

**Fig. 4** Ten-minute $PM_{2.5}$ concentrations colour-coded by retrofit period *(pre-barrier, barrier-on, post-barrier)*.

### 3.6. Key finding

The plastic film barrier did not introduce large external dust peaks inside the apartment, but it doubled the indoor baseline and shifted 45% of residents' $PM_{2.5}$ exposure to periods when natural ventilation was restricted. Once the barrier was removed and controlled airing resumed, the background fell to its lowest observed level.

## 4. Discussion

This study focused solely on fine particulate matter ($PM_{2.5}$). Installing the plastic film barrier cut off natural ventilation: the "quiet" baseline inside the flat rose from ≈ 1.9 µ$g/m^3$ before retrofit to ≈ 3.9 µ$g/m^3$ while the plastic film barrier was in place, then dropped to only 0.8 µ$g/m^3$ after removal. High amplitude peaks (≤ 162 µ$g/m^3$) appeared only during weekday facade work; cooking spikes were lower (≤ 67 µ$g/m^3$) but frequent. Integration of the minute-by-minute excess concentrations shows that, during the barrier period, facade activity delivered 31% of the daily $PM_{2.5}$ dose, cooking 24%, and the enlarged background 45%. The quadratic baseline fitted to background minutes reproduces the data with MAE = 2.3 µ$g/m^3$, < 2% of the largest excursions, so these shares are robust even without an outdoor reference sensor.

## Conclusions

A temporary plastic film barrier can shield occupants from direct construction dust yet simultaneously raise the overall $PM_{2.5}$ background, making routine cooking a significant contributor to the daily dose. Brief, controlled ventilation or a small HEPA purifier would mitigate this side effect. Future work should monitor additional pollutants, add an outdoor reference sensor to quantify infiltration efficiency, and validate the method in multi-room dwellings.